\documentclass[twocolumn]{revtex4}
\usepackage{amssymb}
\usepackage{latexsym}
\usepackage{epsfig}
\usepackage{color}
\usepackage{float}
\usepackage{graphicx}
\usepackage{epstopdf}
\begin{document}

\title{A Tsallisian Universe}
\author{Kavoos Abbasi\footnote{kabbasi@yu.ac.ir}, Shirvan Gharaati\footnote{gharaati@yu.ac.ir}}
\address{Department of Physics, College of Sciences, Yasouj University, Yasouj, 75918-74831, Iran}

\begin{abstract}
In this paper, we adopt the Verlinde hypothesis on the origin of gravity as the consequence of the tendency of systems to increase their entropy, and employ the Tsallis statistics. Thereinafter, modifications to the Newtonian second law of motion, its gravity and radial velocity profile are studied. In addition and in a classical framework, the corresponding cosmology and also its ability in describing the inflationary phases are investigated.
\end{abstract}

\maketitle

\section{Introduction}

The study of relation between thermodynamics and gravity has a long history \cite{gibbs,B1,H1,H2,B2,D1,u1}. In one hand, Gibbs shows that gravitational systems is not extensive \cite{gibbs}, a conclusion in agreement with the Bekenstein entropy of black holes \cite{B1}, which is a non-extensive entropy. On the other, it seems that all gravitational systems satisfy the Bekenstein entropy bound expressed as \cite{ser}

\begin{eqnarray}\label{1}
S_{BE}=\frac{Ac^3}{4G\hbar},
\end{eqnarray}

\noindent where $A=4\pi R^2$ and $R$ denote the area of the system boundary and its radius, respectively, and $k_{B}=1$ (Boltzmann constant). Using this entropy and Clausius relation, one can show that Einstein gravitational field is in fact a thermodynamic equation of state \cite{jacob}. This amazing result is valid in various gravitational and cosmological setups which lead to notable predictions about the behavior of cosmos and gravitational systems \cite{Pad0,Pad1,jacob1,j1,j2,j3,j4,j5,sheyw1,sheyw2,Pad11,msrw,mr,j14,ahepm,rjp,ahepm1,mrea,ahep,bamba,vali}. Motivated by the Gibbs work \cite{gibbs}, non-extensivity of Bekenstein entropy, and based on the Long-range nature of gravity \cite{pla}, recently, the use of non-extensive statistical mechanics (based on possible generalizations of Gibbs entropy) has been proposed to model and study some phenomena such as the cosmic evolution \cite{non13,EPJC,non20,eplT,Tavayef,epjcr,smm,rinf}, black holes \cite{smbh1,smbh2,smbh3,smbh4,smbh5,smbh6,smbh7,mah2019,me,tsallis} and Jeans mass \cite{ob1,ob2}.

In order to find the probable thermodynamic aspects of gravity, Verlinde describes it as the implication of the tendency of systems to increase their entropy \cite{Ver}. An astonishing approach which attracts investigators to itself \cite{Pad2,cai,sheyr,shey,prdc,c1,c2,c3,c4,c5,c6,SMR,MS}. In the framework of generalized entropies, Verlinde hypothesis leads to significant implications on the cosmic evolution \cite{nonK,EN1,ev,mas}, Newtonian gravity \cite{plbmond}, Jeans mass (as a stability criterion) \cite{mnras}, and also gravitational systems \cite{Neto2011,ob3,cjp,Dil2017b,Senay2018b,Kibaroglu2019}. Indeed, differences between generalized entropies and Bekenstein entropy, originated from the non-extensive viewpoint, can $i$) describe the universe inflationary phases \cite{non13,non20,EPJC,rinf}, $ii$) relate Padmanabhan emergent gravity scenario to the Verlinde hypothesis \cite{non13}, and $iii$) propose an origin for MOND theory \cite{plbmond}.

Based on the Verlinde hypothesis \cite{Ver}, the entropy change of a system increases as

\begin{equation}\label{2}
\Delta S=2\pi \frac{mc}{\hbar}\Delta x,
\end{equation}

\noindent when the test mass $m$ has distance $\Delta x=\frac{\hbar}{mc}\equiv\lambda_c$ (reduced Compton wavelength) with respect to the holographic screen (boundary of system). This screen consists of $N$ degrees of freedom calculated by

\begin{equation}\label{3}
N=\frac{Ac^{3}}{G\hbar},
\end{equation}

\noindent in agreement with Eq.~(\ref{1}) and thus $S_{BE}=\frac{N}{4}$ \cite{B1}. Following Refs. \cite{sheyr,shey}, we assume $\Delta x=\eta\lambda_c$ from now, and simple calculations lead to \cite{sheyr,shey}

\begin{equation}\label{4}
F=T\frac{dS}{dA}\frac{\Delta A}{\Delta x}=ma,
\end{equation}

\noindent if $\eta=\frac{1}{8\pi}$ leading to $\Delta x=\frac{\lambda_c}{8\pi}$, as the net force that source $M$ applies to particle $m$, which finally brings it acceleration $a$. We also used the Unruh temperature relation \cite{u1}

\begin{equation}\label{5}
T=\frac{1}{2\pi}\frac{\hbar a}{c},
\end{equation}

\noindent to get Eq.~(\ref{4}). Now, combining $A=4\pi R^2$ and Eq.~(\ref{3}) with

\begin{equation}\label{6}
E=\frac{1}{2}NT=Mc^2,
\end{equation}

\noindent and using Eq.~(\ref{4}), one easily reaches at Newtonian gravity

\begin{equation}\label{7}
a=G\frac{M}{R^{2}}.
\end{equation}

It is also useful to mention that it seems there is a deep connection between generalized entropies and quantum gravity scenarios, and indeed, quantum aspects of gravity may also be considered as another motivation for considering generalized entropies \cite{eplm,barrow}. Tsallis entropy is one of the generalized entropy measures which leads to acceptable results in the cosmological and gravitational setups \cite{tsallis,me,mah2019,non13,Tavayef}. In fact, there are two Tsallis entropies \cite{tsallis,me,mah2019}. One of them has been proposed by Tsallis and Cirto \cite{tsallis} which is confirmed by the multifractal structure of horizon in quantum gravity \cite{barrow} and modifies Eq.~(\ref{1}) as $S\sim A^\delta$ ($\delta$ is a free unknown parameter \cite{eplm}).

The second one has recently been calculated in Ref.~\cite{me} by relying on statistical properties of degrees of freedom distributed on the holographic screen. The result is compatible with a detailed study in the framework of quantum gravity \cite{mah2019}. This case proposes an exponential relation between the horizon entropy and its surface, and we will focus on it in this paper. In the next section, modifications to the Newtonian second law of motion and also Newtonian gravity is derived by using the Tsallis entropy. Its implications on the radial velocity is also addressed. In the third section, after evaluating the Tsallis modification to the gravitational potential, we adopt the approach of paper \cite{kord}, and find out the corresponding Friedmann first equation in a classical way in which a test mass is located on the edge of universe, namely apparent horizon, \cite{kord}. The possibility of obtaining an accelerated universe is also debated in this section. A summary of the work is presented in the last section.
\section{Tsallis Gravity and Dynamics}

Employing Tsallis statistics, it has been recently shown that Eq.~(\ref{1}) is modified as \cite{me}

\begin{eqnarray}\label{11}
&&S_q^T=\frac{1}{1-q}[\exp\big((1-q)S_{BE}\big)-1],
\end{eqnarray}

\noindent in full agreement with quantum gravity calculations \cite{mah2019}. Here, $q$ is a free parameter evaluated from other parts of physics and also observations, and Eq.~(\ref{1}) is recovered when $q=1$ \cite{pla,me,mah2019}. In the non-extensive scenarios, Eq.~(\ref{6}) takes the form \cite{plas,ev}

\begin{eqnarray}\label{12}
E=\frac{1}{5-3q}NT=Mc^2,
\end{eqnarray}

\noindent which approaches Eq.~(\ref{6}) at the appropriate limit of $q=1$.

Now, following the recipe which led to Eq.~(\ref{4}), one can use Eq.~(\ref{11}) to find

\begin{equation}\label{13}
F^T=T\frac{dS_q^T}{dA}\frac{\Delta A}{\Delta x}=ma\exp(\delta\frac{(2+3\delta)Mc^3\pi}{2\hbar a}),
\end{equation}

\noindent where $\delta\equiv1-q$, as the Tsallis second law of motion. Clearly, Eq.~(\ref{4}) is recovered whenever $\delta=0$, and therefore, this approach claims the net force $F^T$ that source $M$ applies to $m$ depends on $M$. In order to obtain the above result, we used $S_{BE}=\frac{N}{4}$ \cite{B1}, and $N=\frac{(5-3q)Mc^2}{T}$. Of course, since the relation $F=ma$ works very well (classical regime), one can deduce that $\delta$ is very close to $0$ meaning that the exponential factor may has non-sensible effects in the classical regime.

The modified form of Eq.~(\ref{7}), called Tsallis gravity, is also obtained as

\begin{equation}\label{14}
a^T=G_q\frac{M}{R^2}\exp(\delta\frac{R_0^2}{R^2}),
\end{equation}

\noindent where $R_0^2\equiv\frac{G\hbar}{c^3\pi}=\frac{l_p^2}{\pi}$, $l_p$ denotes the planck length, and $G_q\equiv\frac{5-3q}{2}G$ in full agreement with \cite{ev}. In order to have a comparison between Tsallis second law of motion and also Tsallis gravity and those of Newton, let us write

\begin{eqnarray}\label{15}
&&\frac{F^T}{F}=\exp(\frac{d}{a}), \nonumber\\
&&\frac{a^T}{a}=(5-3q)\exp(\frac{l}{R^2}),
\end{eqnarray}

\noindent where $d\equiv\delta\frac{(2+3\delta)Mc^3\pi}{2\hbar}$ and $l\equiv\delta R_0^2$. As a crucial point, one should note that, for an event, the sign of $a$ and $a^T$ should be the same (the predictions of different theories about the value of accelerations should address the same motion meaning that both of $a^T$ and $a$ should have the same sign). It leads to this limitation $q<\frac{5}{3}$ meaning that $\delta>-\frac{2}{5}$. Thus, $l$ and $d$ can be negative.

Now, let us compare Eq.~(\ref{14}) with the results of Refs.~\cite{sheyr} and~\cite{shey} where authors employ different entropies in the framework of Verlinde theory, and address two modifications for the Newtonian gravity. Unlike Eq.~(11) of Ref.~\cite{shey}, the modified gravity obtained in Ref.~\cite{sheyr} (Eq.~(17)), diverges at large distances ($R\gg1$). Of course, both of them claim that gravitational force between source $M$ and test particle $m$ can vanish for some points on their interface line, a property incompatible with the Newtonian gravity and experience. From Eq.~(\ref{14}), one can easily see that the obtained gravitational force does not diverge at large distances where it will be ignorable. Thus, it seems that this equation is a more reliable modification to Newtonian gravity compared with those of Refs.~\cite{sheyr} and~\cite{shey}.
\subsection*{Velocity profile}

For a circular motion at radius $r$ with velocity $v$, and thus acceleration $\frac{v^2}{r}(\equiv a^T)$, obeying Eq.~(\ref{14}), one reaches

\begin{eqnarray}\label{16}
v=\sqrt{\frac{G_q}{r}}\exp(\frac{l}{2r^2}),
\end{eqnarray}

\noindent which implies that we should have $q<\frac{5}{3}$ to get real values of velocity.

On the other hand, if one assumes the mass $m$ in the gravitational field of source $M$ feels the force $GMm/r^2$, then using~(\ref{13}), we can write

\begin{eqnarray}\label{17}
GMm/r^2=F^T,
\end{eqnarray}

\noindent yields

\begin{eqnarray}\label{18}
GM/r=v^2\exp(\frac{d r}{v^2}),
\end{eqnarray}

\noindent for $a\equiv\frac{v^2}{r}$, finally leading to

\begin{eqnarray}\label{19}
v^2\simeq \frac{GM}{r}-d r,
\end{eqnarray}

\noindent if we expand $\exp(\frac{d r}{v^2})$ as $1+\frac{d r}{v^2}$. For a constant $d$, this approximation is valid when radial acceleration ($\frac{v^2}{r}$) is small. Indeed, in this manner, the $dr$ term leads to increase in the velocity of particle $m$, compared with the Newtonian case for which $v^2\simeq \frac{GM}{r}$, if $d<0$.

\section{A Tsallis cosmology}

In order to find the Friedmann first equation corresponding to the obtained Tsallis gravity, we follow the classical viewpoint fully described in Ref.~\cite{kord}. Tsallis gravitational potential can be easily calculated by using Eq.~(\ref{14}) as

\begin{eqnarray}\label{21}
\phi(r)=-\frac{G_qM}{r}\Sigma_{n=0}^{\infty}\frac{l^n}{n!(2n+1)r^{2n}},
\end{eqnarray}

\noindent where we used $\exp(\frac{l}{r^2})=\Sigma_{n=0}^{\infty}\frac{l^n}{n!r^{2n}}$ to reach the above result. Considering a test particle on the edge of a flat FRW universe, and following the recipe of Ref.~\cite{kord}, this equation leads to

\begin{eqnarray}\label{22}
H^2=\frac{8\pi G_q}{3}\rho\Sigma_{n=0}^{\infty}\frac{l^nH^{2n}}{n!(2n+1)},
\end{eqnarray}

\noindent in which $\rho$ is the cosmic fluid density, $H$ denotes the Hubble parameter, and we used the fact that apparent horizon is located at $r=\frac{1}{H}$. Moreover, the standard Friedmann first equation \cite{kord} is recovered at the desired limit of $q=1$ (or equally, $\delta=0(\parallel l=0)$).

\subsection*{Accelerated universe}

Bearing the fact that Hubble parameter decreases during the cosmic evolution in mind, rewriting Eq.~(\ref{22}) as

\begin{eqnarray}\label{23}
\frac{H^2}{\Sigma_{n=0}^{\infty}\frac{l^nH^{2n}}{n!(2n+1)}}=\frac{8\pi G_q}{3}\rho,
\end{eqnarray}

\noindent and keeping terms up to the $H^4$ term in LHS (the first corrective term to the standard cosmology ($H^2=\frac{8\pi G}{3}\rho$) due to Tsallis gravitational potential), one easily reaches at

\begin{eqnarray}\label{24}
H^2\simeq\frac{3}{2l}\big(1\pm\sqrt{1-\frac{32\pi G_ql}{9}\rho}\big).
\end{eqnarray}

In order to have real solutions for $H^2$, this equation claims that there is a maximum bound on density of cosmic fluid as $\rho_{max}=\frac{9}{32\pi G_ql}$ at which universe feels a de-Sitter phase with $H=\sqrt{\frac{3}{2l}}$ when $l>0$. As universe expands, $\rho$ decreases and when $\rho=0$, the positive branch experiences again the primary de-Sitter phase ($H=\sqrt{\frac{3}{l}}$ for $l>0$) but forever, while the universe expansion rate vanishes for the negative solution. In fact, the vacuum solution ($\rho=0$) of the above Friedmann first equation is an inflationary universe for positive branch, and a Minkowski universe for the negative branch.

\section{Summary}

In the framework of Verlinde hypothesis on the origin of gravity, we employed the recently proposed Tsallis entropy \cite{mah2019,me} to find its implications on the Newtonian dynamics (second law of motion) and gravity. The velocity profile in a circular motion has also been analyzed. Finally, adopting the classical approach to get the Friedmann first equation described in Ref.~\cite{kord}, the corresponding cosmology was achieved after finding the Tsallis gravitational potential. The obtained modified Friedmann first equation~(\ref{23}) includes a complex function of $H$.

Since Hubble parameter decreases during the cosmic evolution, and because the standard Friedmann first equation ($H^2=\frac{8\pi G}{3}\rho$) has notable achievements, we only focused on the first corrective term due to the Tsallis gravitational potential (i.e. we only hold terms up to $H^4$ in writing Eq.~(\ref{24})). We saw that, in some situations and depending on the value of $\delta$, the resulting equation addresses the (anti) de-Sitter universes with $H=\sqrt{\frac{3\pi}{2\delta l_p^2}}$ and $H=\sqrt{\frac{3\pi}{\delta l_p^2}}$, depending on $\rho$. It also admits an upper bound on energy density of cosmic fluid of order of $\frac{l_p^{-2}G^{-1}}{(2+3\delta)\delta}\sim\frac{10^{81}}{(2+3\delta)\delta}$. We also obtained that there are two branches for the assumed approximation. Whenever $\rho=0$, the positive branch, depending on the value of $\delta$, guides us to an eternal (anti) de-Sitter phase, and the negative branch addresses a Minkowskian fate for universe.

\end{document}